\documentclass[twocolumn,twocolappendix]{aastex631}
\usepackage{apjfonts}    
\usepackage{natbib}      
\usepackage{amsmath}
\usepackage{xcolor}      
\usepackage{booktabs}
\usepackage{tabularx}

\usepackage{booktabs}

\usepackage{amsmath,tabularx} 
\usepackage{graphicx}
\newcommand{\beq}{\begin{equation}}
\newcommand{\beqa}{\begin{eqnarray}}
\newcommand{\eeq}{\end{equation}}
\newcommand{\eeqa}{\end{eqnarray}}
\newcommand{\ba}{\[\begin{aligned}}
\newcommand{\ea}{\end{aligned}\]}

\shorttitle{}
\shortauthors{}

\providecommand{\nolinenumbers}{}
\AtBeginDocument{\nolinenumbers}

\begin{document}

\title{
Hybrid Strategy for Coordinated Interstellar Signaling: Linking the Galactic Center and Extragalactic Bursts
\\
}


\author{Naoki Seto}
\affil{Department of Physics, Kyoto University, Kyoto 606-8502, Japan}
\email{seto@tap.scphys.kyoto-u.ac.jp}

\begin{abstract}
The search for extraterrestrial intelligence (SETI) is largely limited by the vastness of the signaling parameter space. The concurrent signaling scheme offers a framework in which civilizations can coordinate their transmission and reception by referring to a common astrophysical event. Building on this idea, I propose a hybrid strategy that combines the Galactic Center as a spatial reference with an extragalactic burst as a temporal marker. If such a scheme is indeed employed, the sky area to be surveyed in SETI could be reduced by more than two orders of magnitude, based solely on existing astronomical data. I examine records of three types of extragalactic bursts (supernovae, neutron star mergers, and gamma-ray bursts [GRBs]) to identify suitable  temporal markers. Among them, GRB 221009A is particularly notable  due to its high fluence and favorable sky location. 

\end{abstract}





\section{introduction}
Searches for intentional signals from extraterrestrial intelligence (ETI) have been conducted for more than sixty years~\citep{drake1961ozma,Tarter2001,2013ApJ...767...94S,2021lcfb.book.....L}, yet no definitive detection has been achieved. 
While it is possible that intelligent civilizations are exceedingly rare in our Galaxy, it is also conceivable that such signals do exist but  have not yet been detected. 
The relevant parameter space (including frequency, direction, and time) is vast and multidimensional~\citep{Horowitz1993}. 
Given the limited available resources (both observational and computational), performing a comprehensive and high-resolution search across all dimensions remains a formidable challenge. 
Strategic narrowing of this parameter space would be essential for practical SETI efforts~\citep{Shostak2011,Wright2018}.

These observational challenges on the receiver's side are likely foreseeable by the transmitter as well. 
Therefore, even in the absence of prior communication, some form of implicit adjustment may emerge to reduce the burden of search and transmission for both sides.
Such spontaneous alignment within a strategic space is known in game theory as a ``Schelling point,'' where players tend to converge on a particular choice without prior coordination (for example, based on symmetry, uniqueness, or conspicuousness)~\citep{Schelling1960}. 
In the context of SETI, several previous studies have explored the potential relevance of Schelling points for mitigating the difficulties of searching across a vast parameter space~\citep{1959Natur.184..844C,1975Natur.254..400P,1977Icar...32..464M,1980Icar...41..178M,1994Ap&SS.214..209L,Corbet:1999vv,Shostak2009,Wright2020}.

Building on this idea, I have proposed a strategy, termed the concurrent signaling scheme~\citep{Seto2019,Seto2021,Seto2024}. 
In this approach, both Galactic transmitters and receivers refer to a common astronomical anchor event within the Milky Way to coordinate the direction of signaling over time. 
This method can substantially reduce the effective search space, as it allows for distance-independent signal reception (from a given direction), in the sense that all ETIs within a certain range can be searched collectively. 
This concurrent signaling scheme  thus provides a viable means of optimizing SETI strategies under resource limitations.

To realize this scheme, the Galactic anchor event must be conspicuous and have a well-defined time of occurrence. 
In addition, its three-dimensional position must be determinable with sufficient accuracy by both the transmitter and the receiver independently. 
The anchor can be either a past or a future event~\citep{Seto2019,Seto2021}.

I have examined several possible candidates for  Galactic anchor events, including future Galactic neutron star binary (NSB) mergers, past Galactic supernovae recorded within the last $\sim2000$ years, and pericenter passages of stars near the central black hole Sagittarius A*. 
However, each of these options poses practical challenges. 
NSBs  with sufficiently short orbital periods have yet to be detected. 
Most supernovae have large uncertainties in their distances~(see, for example,  \citealt{Kaplan:2008qm}). 
The vicinity of Sagittarius A* is densely populated with stars \citep{Genzel2010}, making it difficult to identify a limited number of Schelling-like anchors.

To address these challenges,  I propose a novel hybrid strategy that combines two types of references. 
One is the Galactic Center, which would serve as a positional reference point applicable throughout the entire Galaxy with very high precision~\citep{2021A&A...647A..59G}.
The other is a conspicuous extragalactic burst that serves as a temporal and directional marker. 
When the burst is sufficiently distant, the uncertainty in its distance becomes negligible for determining the signal search direction. 
This combination enables a substantial reduction of the target sky area for SETI observations.

To evaluate the feasibility of our method, we consider three types of burst phenomena as potential extragalactic temporal markers: supernovae, NSB mergers, and gamma-ray bursts (GRBs) (see also \citealt{Corbet:1999vv}). 
For each category, we examine the suitability of well-documented prominent events for our extragalactic bursts.

This paper is organized as follows.
In Section 2, we formulate the core idea of the concurrent signaling scheme.
In Section 3, we review previously proposed Galactic reference anchors and discuss their limitations.
In Section 4, we introduce the hybrid scheme and derive the formula for computing the search direction as a function of time.
In Section 5, we present a geometrical analysis of the scheme, taking into account causal relationships.
We also evaluate the uncertainties in the search direction arising from estimation errors in the distance to the Galactic Center.
In Section 6, we explore three types of extragalactic bursts as temporal markers and provide case studies based on existing astronomical records.
In Section 7, we discuss several additional aspects of the proposed strategy.
Finally, in Section 8, we briefly summarize this study.

\section{concurrent signaling scheme}

\subsection{Basic Idea}
The concurrent signaling scheme is a potential method for coordinating efficient interstellar signal transmission and reception without prior communication between the involved civilizations~\citep{Seto2019}.  
In this scheme, it is assumed that they commonly select a conspicuous astronomical event whose distance and epoch can be estimated with reasonable accuracy by both parties.

\begin{figure}[t]
 \begin{center}
  \includegraphics[width=8cm,clip]{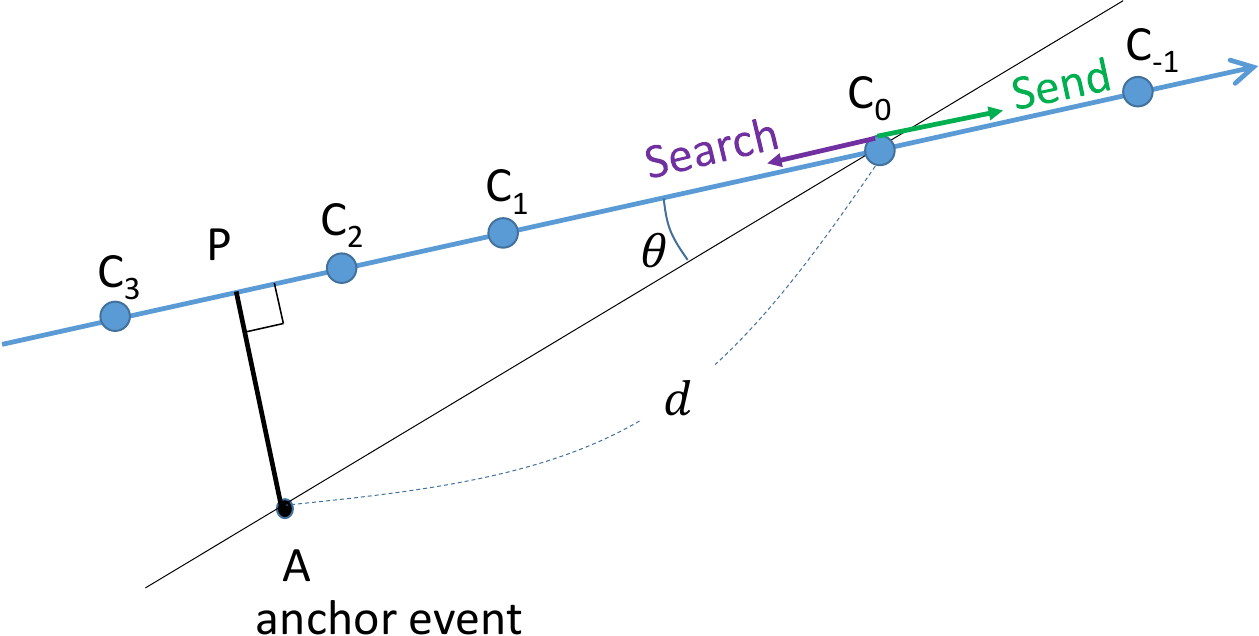}
   \caption{Schematic illustration of the signaling geometry.  
A set of civilizations ($C_{-1}, C_0, C_1, \dots$) are distributed along an oriented line.  
An astronomical event occurs at point $A$, with its closest point on the line denoted by $P$.  
At the location of civilization $C_0$, the distance to the event is $d$, and the angular offset from the line is $\theta$.  
Using the concurrent signaling scheme, the civilizations can coordinate their signaling activities without prior communication,  
relying solely on shared knowledge of event $A$.  }
  \label{figure:fig1}
 \end{center}
\end{figure}

Let us consider civilizations \( C_i \) located along a given oriented line (see Fig.~\ref{figure:fig1}).  
In this scheme, transmitters emit signals synchronized with a hypothetical signal passing through a specific point on the line at a specific time, thereby forming a temporally coherent group of intentional signals that propagate along the line.  
An important aspect of this scheme is that the reference point and the associated epoch on the line can be commonly determined based on a shared reference to the anchor event $A$.

In Figure~\ref{figure:fig1}, for a given line and an external anchor $A$, the point of closest approach \( P \) will be uniquely selected as the Schelling point on the line.  
Regarding the reference time at point \( P \), there are two natural candidates:  
(i) the epoch \( t_A \) when the anchor event occurs (on a simple time slice), and  
(ii) the epoch \( t_A + {AP}/c \) when the electromagnetic signals from the anchor event reach point \( P \).  
To treat both cases in a unified manner, we represent  the reference time as \( t_A + q \times {AP}/c \) with a parameter \( q \), where \( q = 0 \) corresponds to case (i) and \( q = 1 \) to case (ii).  
This parameter effectively shifts the reference time by the baseline interval \( {AP}/c \), which is independent of the location along the line~\citep{Seto2021}.

In this scheme, the search directions are simply antipodal to the transmission directions (see Fig.~\ref{figure:fig1}).  
Therefore, it is not necessary to treat transmission and reception separately, and we hereafter take the perspective of a searcher.

{In the original proposal by \cite{Seto2019}, I fixed the signaling schedule such that the artificial signals were synchronized with the reference anchor signal at an infinite distance. To simplify the geometrical interpretation and generalize the scheme, in the subsequent paper \citep{Seto2021}, I reformulated the scheme using the closest approach point $P$ and introduced an additional parameter $q$. The  scheme in \cite{Seto2019} is recovered when $q = 0$.
}

Let us assume that we are located at the point \( C_0 \) in Fig.~\ref{figure:fig1}.  
Using only the information from the anchor event $A$ (with its distance \( d \) and angle \( \theta \) as defined in Fig.~\ref{figure:fig1}), we can determine when to search for ETI signals along the blue oriented line.  
The anchor event is observed at the epoch \( t_A + d/c \), while the intentional signals from civilizations along the line arrive at \( t_A + (\cos\theta + q\sin\theta)\, d/c \).  
The time difference between the two arrivals is given by
\beq
\Delta t = (\cos\theta + q\sin\theta - 1)\, \frac{d}{c}.  \label{co1}
\eeq
{In terms of the nondimensional time $x\equiv \Delta t/(d/c)$, we have
\beq
x = \cos\theta + q\sin\theta - 1.  \label{co2}
\eeq
}
It is important to notice that, in general, the intentional signals are not synchronized (namely \( \Delta t = 0 \)) with the anchor signal (except for special angles, e.g., \( \theta = 0 \)).

\begin{figure}[t]
 \begin{center}
  \includegraphics[width=8cm,clip]{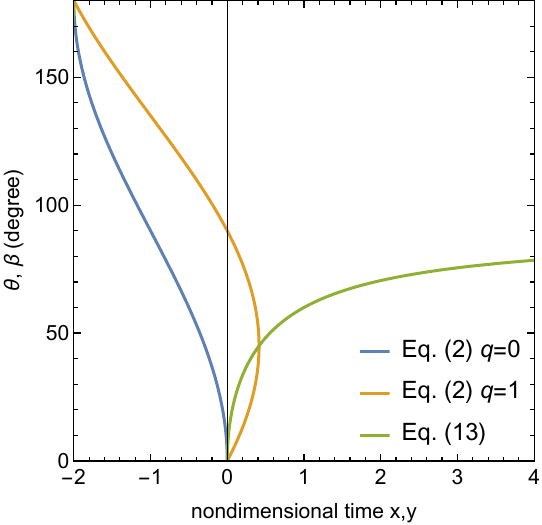}
   \caption{Target offset angles given  in Eqs. (\ref{co2}) and (\ref{tt4}) from the burst direction as a function of the nondimensional times $x$ and $y$.  For $q=1$, we have a solution for $\theta$ only with $-2\le x \le \sqrt{2} - 1$. }
  \label{figure:tt}
 \end{center}
\end{figure}

So far, we have focused on signaling along a single oriented line, but the framework can be extended to an ensemble of lines distributed arbitrarily in three-dimensional space.  
Owing to the symmetry of the setup, the search directions at a given time and location form a ring on the sky, centered on the axis connecting the observer and the anchor event.

In fact, we can invert Eqs.~(\ref{co1}) or (\ref{co2}) to solve for the offset angle \( \theta \) of the search ring as a function of the time difference \( \Delta t \) or the normalized one $x$ (see Fig. \ref{figure:tt} for $q=0$ and 1).  
In the case of \( q = 0 \), we have a solution \( \theta \in [0, \pi] \) only in the time range \( \Delta t \in [-2d/c, 0] \). In this case, the signal must be searched for before the anchor event is actually observed, requiring an advance prediction of the event ~\citep{Seto2019}.  On the other hand, for \( q = 1 \), we have two solutions \( \theta \in [0, \pi/2] \) for  \( \Delta t \in [0,\, (\sqrt{2}-1)\, d/c] \), so that the intentional signal arrives after the observation of the event.
This post-observation ($q = 1$) approach enables the use of historical astronomical records  to select anchor events~\citep{Seto2021}.

{In Fig.~1, relative to the oriented signaling line shown in blue, the point $C_3$ is located upstream of the point of closest approach $P$. Its transmission (and reception) angle angle is
\[
\theta = \pi - \angle PC_3A > \pi/2,
\]
and, even for $q = 1$, transmission must be done before the observation of the anchor event (i.e., $x < 0$), as illustrated in Fig.~2.}

Note that, without the timing error for \( \Delta t \), the sharpness of the search ring \( \delta \theta \) is determined by the estimation error in the distance \( d \) to the anchor event~\citep{Seto2019}.

Each civilization performs the same procedure independently at its own location, thereby substantially narrowing the search directions at any given time.  
In this way, the concurrent signaling scheme efficiently links transmitters and searchers without any prior coordination, relying solely on the shared reference to a conspicuous external event.  
Therefore, this scheme can be regarded as a promising candidate for a Schelling point in the space of signaling strategies.

{In the transmission scheme known as the SETI ellipsoid \citep{1977Icar...32..464M,1980Icar...41..178M,1994Ap&SS.214..209L}, a civilization is assumed to transmit artificial signals in all directions at the arrival epoch of a reference burst. This requirement can be demanding, particularly in terms of energy storage and transmission capacity. Moreover, the searcher must examine a two-dimensional ellipsoidal surface spanning the full $4\pi$ steradians of the sky. In contrast, the concurrent signaling scheme (including the hybrid version proposed in Section 4) requires only ring-like sending directions on the sky at each epoch.  In addition, as noted earlier, this scheme allows searches that are independent of the distances to the transmitters.}

\section{Proposed Galactic Anchors}

In this section, I briefly introduce three types of potential Galactic anchors that I have proposed previously.

\subsection{Future Neutron Star Binary Mergers}

In my first paper \citep{Seto2019}, I discussed the possibility of using a future merger of a Galactic NSB as a signaling anchor with \( q = 0 \) (see also \citealt{2018ApJ...862L..21N}). Written shortly after the detection of GW170817 \citep{LIGOScientific:2017vwq}, the study assumed a relatively high Galactic NSB merger rate of \( 1.5 \times 10^{-4}~{\rm yr^{-1}} \) \citep{Kyutoku:2018aai}.

Approximately \( 10^4 \) years prior to merger, an NSB emits gravitational wave (GW) at a frequency around 5~mHz, near the optimal sensitivity band of the Laser Interferometer Space Antenna (LISA), which is scheduled for launch around 2035 \citep{LISA:2022yao}. By analyzing the GW, it would be possible to determine the distance \( d \) and merger epoch of the binary with high precision (e.g., \( \delta d/d \sim 0.01 \)), based on first-principles physics \citep{schutz1986,Kyutoku:2018aai}.

However, at present, there is no identified NSB system suitable for use as a reference. Furthermore, since the landmark detection of GW170817 in 2017, the LIGO-Virgo-KAGRA (LVK) network has experienced a notable scarcity of confident NSB merger detections \citep{KAGRA:2021duu}. Consequently, estimates of the Galactic NSB merger rate have decreased over the past \(\sim 6\) years.

\subsection{Historical Supernovae}

In the second paper \citep{Seto2021}, I proposed leveraging historical supernovae (SNe) recorded in the past \(\sim 2000\) years \citep{Stephenson2002} as anchor events with \( q = 1 \). I selected six SNe with well-identified remnants for our scheme. While the observed epochs of these SN explosions  can be specified with sufficient precision, the estimated distances to the remnants carry large uncertainties compared to robust techniques based on binary orbital dynamics. For example, the distance to the Crab Pulsar (SN1054) has an uncertainty of about 20\% \citep{Kaplan:2008qm}. As a result, the corresponding search rings generally become broad, limiting the ability to efficiently narrow down the survey directions (see also \citealt{2023AJ....166...79N}, which includes an analysis with SN1987A).

\subsection{Pericenter Passages of Stars around the Galactic Center}

The Galactic Center is a salient place in our Galaxy, and in fact, several SETI programs have already been conducted in its direction \citep{1985AcAau..12..369S,2017AcAau.139...98W,2021AJ....162...33G,2022PASA...39....8T,2023AJ....165..255S}. Despite its remoteness, the distance to the Galactic Center has been determined with exceptional precision \citep{2021A&A...647A..59G}:
\begin{equation}
r = 8275 \pm 9_{\rm stat} \pm 33_{\rm sys}~{\rm pc,}
\label{dgc}
\end{equation}
corresponding to a relative uncertainty of approximately 0.5\% level ($\Delta r / r \sim 5 \times 10^{-3}$).
This high accuracy arises from the large mass of the central black hole, which induces substantial variations in both radial velocities and astrometric positions for nearby stars. Hereafter, the symbol \( r \) represents the distance to the Galactic Center. 

Unfortunately,  in the vicinity of the central black hole, we know of no prominent explosive events suitable for setting timing references. As an alternative, my third paper \citep{Seto2024} proposed using the pericenter passages of the prominent B-type star S2 as timing references.  Since the orbital motion is expected to  exhibit high regularity over timescales of about a century, it is possible to compile a list of past and future pericenter passages. By applying Eq.~(\ref{co1}) with both \( q = 0 \) and \( q = 1 \) to this list, one can repeatedly survey civilizations in the directions around the central black hole at intervals of 16.05 years (the orbital period of S2).

However, due to the large number of stars near the central black hole \citep{Genzel2010}, there remains considerable arbitrariness in selecting a star   to define the reference epochs.

\section{Hybrid Approach}
We now turn to our new approach.
\subsection{Scheme Selection}
Given the limitations of purely Galactic anchors, I first explored various hybrid schemes that combine a prominent extragalactic burst with the Galactic Center. The underlying idea is to leverage the precisely known distance to the Galactic Center, while introducing an unambiguous timestamp via the extragalactic burst, since the Galactic Center itself lacks conspicuous burst-like events.

{
I subsequently evaluated the candidate hybrid schemes without being constrained by the trajectory of my earlier studies. 
I then selected the simplest and most symmetrical scheme from the perspective of a Schelling point. This scheme also avoids an undesirable latency of order ${O}(r/c)$ when initiating the search, at least in the burst direction, thereby allowing immediate application.\footnote{For example, as possible extensions of the configuration shown in Fig.~1, one could fix the signal transition epoch at the closest approach $P$ (to the Galactic Center $A$) when the burst reaches either $P$ or $G$. The former choice breaks the symmetry of the target sky directions, whereas the latter introduces an undesirable time delay of ${O}(r/c)$ after the burst signal is observed.}}
In the remainder of this paper, we focus on this selected scheme.

In this scheme, we use the wavefront of the burst signal (propagating at the speed of light), specifically when it reaches the Galactic Center as illustrated in Fig. 3. 
{By construction, this wavefront is uniquely determined and does not depend on the Galactic position of a civilization.
}
For a given oriented line (shown in blue), the reference point and epoch for coordinating signals are set by the outward intersection $K$, which is independent of the civilizations located on the line. Along this blue line, the coherent intentional signals are synchronized with the component of the burst signal that is scattered once at point $K$.

{
Throughout this paper, I use the term “anchor” to refer to Galactic events that provide both spatial and temporal information, as in my previous studies (see Sections 2 and 3). In contrast, in the newly introduced hybrid scheme, I    explicitly refer to the Galactic Center as the spatial reference and to extragalactic bursts as temporal markers,  without using the term “anchor.” This distinction is intended to avoid confusion between the previous and present approaches.}

 \subsection{Geometrical Set-up}

We adopt the Galactic coordinate system defined by the Galactic longitude \( l \) and latitude \( b \). The Galactic plane corresponds to \( b = 0 \), and the  Galactic Center $G$ is located at \( (l, b) = (0, 0) \), with its  distance $r$ given in Eq.~(\ref{dgc}).

In the corresponding Cartesian coordinate system, the unit vector pointing in the direction \( (l, b) \) is 
\[
(\cos b \cos l, \cos b \sin l, \sin b),
\]
with  \( (1, 0, 0) \) pointing toward the Galactic Center.  
Let \((l_B,b_B)\) denote the direction of the reference extragalactic burst, and let \(R\) be its distance from the Sun.
In Fig.~3, the angle \(\theta_B\) between the Galactic Center and the burst is
\begin{equation}
\cos\theta_B = \cos b_B \cos l_B. \label{tb0}
\end{equation}
We define $ K_B$ as the intersection between the line $SB$ and the specific burst wavefront.

\begin{figure}[t]
  \centering
  \includegraphics[width=8cm,clip]{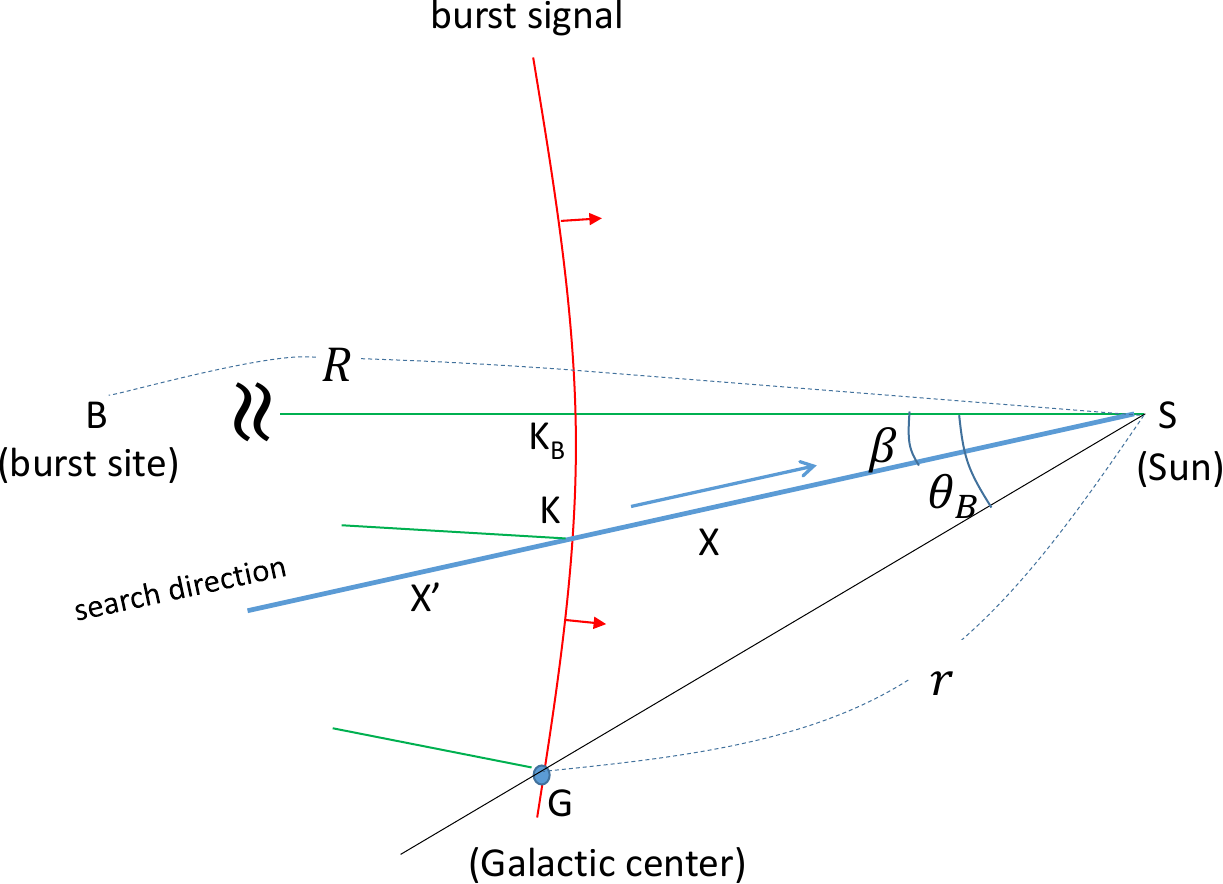}
 \caption{
Schematic illustration of the hybrid scheme.  
The red curve represents the wavefront of the burst signal at the moment it reaches the Galactic Center $G$, where ${ BG} = { BK} = { BK}_B$.  
For the blue signaling line, the reference point and epoch are defined by the outward intersection $K$ with the wavefront.  
At a given time after the burst arrival, the search directions are symmetric around the burst direction.
}
  \label{figure:fig2}
\end{figure}

\subsection{Search Direction}
In Fig. 3, let us  consider the reception of intentional signals along the blue line at the position of the Sun, $S$. 
We evaluate the time delay $\tau$ of the reception  after the burst arrival, based on the corresponding path length difference. Here we assume a flat spacetime and ignore the motions of the involved civilizations relative to the Galactic Center. The aberration effect will be discussed later (see also \citealt{Seto2024} for the effects  of the gravitational potential).

In Fig. 3, the reference burst surface is axially symmetric about the line  $SB$.  
For a civilization located outside this surface, such as at the position of the Sun \(S\), the corresponding search directions form a ring centered on the burst direction. We denote the opening angle of the ring by $\beta$. 
{As shown below, at each point (e.g., $S$ in Fig.~3), the angular size $\beta$ is determined from the positional information of the Galactic Center and expressed as a function of the time $\tau$ after  the burst
is observed }

Applying the law of cosines to triangle \(\triangle {SGB}\), we obtain the distance from the burst to the Galactic Center as
\beq
{BG} = \sqrt{R^2 + r^2 - 2 R r \cos \theta_B}, \label{bg}
\eeq
where \(R\) is the distance to the burst.  
We also have \({BG} ={BK} = {BK_B}\).

We now consider triangle \(\triangle {SKB}\).  
Applying the law of cosines, we obtain
\beq
{BK}^2 ={BG}^2 = R^2 + {KS}^2 - 2 R \cdot {KS} \cos \beta.
\eeq
Then, the signal path length $KS$ is given by
\begin{equation}
KS = R \cos \beta - \left[ R^2 \cos^2 \beta - (R^2 - {BG}^2) \right]^{1/2}.
\label{ks}
\end{equation}
{
The search direction is tangent to the red surface at an angle $\beta$ satisfying
\begin{equation}
R^2 \cos^2 \beta - \left(R^2 - {BG}^2\right) = 0.
\label{eq:tangent-condition}
\end{equation}
}

The distance between \({K_B}\) and the Sun is
\beq
{K_BS} = R -{BK_B} = R - {BG}. \label{kbs}
\eeq
The time delay \(\tau\) is then defined as
\beq
\tau = \frac{{KS} -{K_BS} }{c}. \label{bast}
\eeq
Using Eqs.~(\ref{bg}), (\ref{ks}), and (\ref{kbs}), we can express the time delay \(\tau\) in terms of \(R\), \(\theta_B\), and \(\beta\).  
Conversely, the ring opening angle \(\beta\) can be determined as a function of the time delay \(\tau\).

\subsection{Plane Wave Approximation}

In the distant limit \( r / R \to 0 \), the reference burst surface becomes approximately planar and perpendicular to the burst direction.  
In this case, we can apply the plane wave approximation:
\[
{K_BS} = r \cos \theta_B, \quad
{KS} = \frac{ r \cos \theta_B }{ \cos \beta },
\]
and the corresponding delay time becomes
\beq
\tau = \frac{r}{c} \cos \theta_B \left( \frac{1}{\cos \beta} - 1 \right). \label{tt2}
\eeq
{
We can  solve
\begin{equation}
\beta = \arccos\!\left[\frac{r \cos\theta_B}{c\,\tau + r \cos\theta_B}\right],
\label{tt3}
\end{equation}
which, in terms of the normalized time $y \equiv \tau/(r \cos\theta_B / c)$,  reduces to
\begin{equation}
\beta = \arccos\!\left[\frac{1}{y + 1}\right].
\label{tt4}
\end{equation}
These expressions provide the leading-order approximations for distant bursts. Equation~(\ref{tt4}) is presented in Fig.~2. We have $\beta = 0$ at $y = 0$ and $\beta \to \pi/2$ as $y \to +\infty$.
}

Including first-order corrections in \( r/R \), the delay time can be expanded as
\begin{align}
\tau
&= \frac{r}{c} \cos \theta_B \left( \frac{1}{\cos \beta} - 1 \right) \notag \\
&\quad + \left( \frac{r}{R} \right) \frac{r}{2c} \left( \sin^2 \theta_B + \frac{ \cos 2\theta_B - \cos 2\beta }{2 \cos^3 \beta } \right) \notag \\
&\quad + \mathcal{O}\left[ \left( \frac{r}{R} \right)^2 \right]. \label{t2}
\end{align}
{
In subsection~5.1, we show that, with our current precision of \( \Delta r / r \sim 0.005 \), the second term becomes negligible for distant bursts at \( R \gtrsim 2~\mathrm{Mpc} \). As shown in Eq.~(\ref{tt2}), the scheme then becomes effectively independent of \( R \).
}

\section{Geometric Properties of the Hybrid Approach}
In this section, we discuss basic geometrical properties of the hybrid scheme.
\subsection{	Causal  Constraints on Search Regions}
Let us consider the propagation of intentional signals along the blue line shown in Fig.~3.  
In this scheme, a civilization at point $K$ transmits its intentional signal in $2\pi$ sky directions outward from the burst sphere when the burst reaches $K$. 
{The situation at point \( K \) is essentially the same as the transmission scheme in the SETI Ellipsoid \citep{1977Icar...32..464M,1980Icar...41..178M,1994Ap&SS.214..209L}. Civilizations located downstream of \( K \) (such as points \( S \) and \( X \)) can transmit or receive signals after observing the burst, with the delay time depending on the sky direction and formally given by Eq.~(\ref{bast}) (evaluated perturbatively in subsection~4.4).
}  
In contrast, a civilization located upstream of $K$ (such as point $X'$) must act before observing the burst in order to satisfy the temporal coherence along the line.

This causality constraint can be confirmed geometrically using the triangle inequality for triangle \(\triangle {KBX'}\):
\[
{KB} - {KX'} < {BX'}.
\]

It would be challenging, even for highly advanced Galactic civilizations, to predict the arrival time of a distant burst like GRB~221009A \citep{2023ApJ...946L..31B} more than \(\sim 10\) years in advance.\footnote{The situation is different for an NSB merger, as commented later.}  
Therefore, we may reasonably assume that intentional signals are transmitted only after the burst is actually observed at each civilization.
This implies that participating civilizations must lie downstream of the wavefront  shown in red in Fig.~\ref{figure:fig2}.

We now consider the situation at the Sun, using the plane wave approximation given in Eq.~(\ref{tt2}).  
Under this approximation, the downstream condition corresponds to
\begin{equation}
0 \le \theta_B < 90^\circ. \label{theta_downstream}
\end{equation}
Note that the angle \(\theta_B\) depends on the location of a civilization within the Galaxy.

{
Using a large number of extragalactic bursts, we can in principle search for civilizations throughout our Galaxy, except for the regions near the two endpoint directions along the line connecting our position \( S \) to the Galactic Center \( G \) in Fig.~3; the segment \( SG \) between these points remains accessible.
}

Extragalactic burst records date back only to 1885 \citep{1885Natur..32..499C}.
Accordingly, for the immediate application of the hybrid scheme, the delay time \(\tau\) of a reference burst must satisfy  
\begin{equation}
\tau \lesssim 100~\mathrm{yr} \quad (\ll r / c \sim 2.7 \times 10^4~\mathrm{yr}). \label{tau_max}
\end{equation}

From Eq.~(\ref{tt2}), this constraint leads to
\begin{equation}
0 \le \beta \ll 1 \quad \text{or} \quad 0 \le \cos \theta_B \ll 1. \label{beta_small}
\end{equation}

\begin{figure}[t]
 \centering
 \includegraphics[width=8cm]{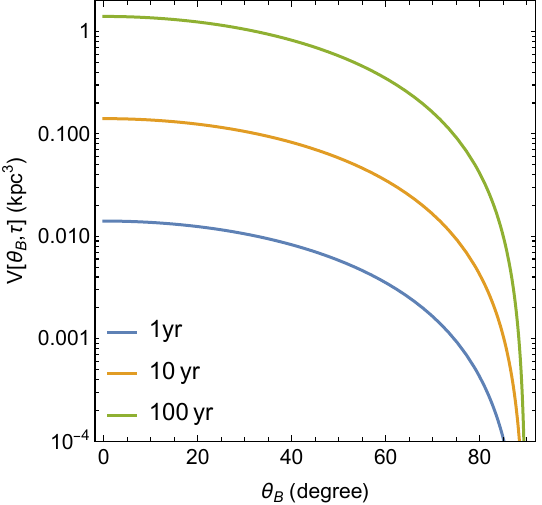}
 \caption{
Accessible volume \(V(\theta_B, \tau)\) as a function of \(\theta_B\), shown for elapsed times \(\tau = 1, 10\), and \(100~\mathrm{yr}\). 
Note that \(V(\theta_B, \tau)\) is approximately proportional to \(\tau\).}
 \label{figure:fig3}
\end{figure}

We can also evaluate the volume of the region accessible within the time interval \([0, \tau]\).
 The  region is a cone with  height $r\cos\theta_B$ and  opening angle $\beta$, and we obtain its volume as
\begin{equation}
V(\theta_B,\tau) = \frac{(r \cos \theta_B)^3 \tau \left( \tau + 2r \cos \theta_B / c \right)}{3 \left( \tau + r \cos \theta_B / c \right)^2},
\end{equation}
where we applied Eq. (\ref{tt2}).
Numerical results for \(\tau = 1\), 10, and 100~yr are presented in Fig.~\ref{figure:fig3}.  
To avoid significant loss in  accessible volume, we require
\begin{equation}
0 \le \theta_B \lesssim 80^\circ. \label{theta_volume}
\end{equation}

Compiling inequalities~(\ref{theta_downstream}), (\ref{beta_small}) and (\ref{theta_volume}),  our survey geometry should  satisfy
\begin{equation}
0 \le \theta_B \lesssim 80^\circ,~~\beta \ll 1.\label{beta1}
\end{equation}

In the limit \(\tau \ll r \cos \theta_B / c\), the accessible volume reduces to
\[
V(\theta_B, \tau) \sim \frac{(r \cos \theta_B)^2 c \tau}{3},
\]
which is approximately proportional to the elapsed time \(\tau\).

\subsection{Preferred Galactic Coordinates of  a Burst}

Galactic stars are highly concentrated near the  mid-plane of the disk at \(b = 0\).  
Therefore, a larger number of stellar systems are contained in the accessible volume associated with a burst at low Galactic latitude.  
We thus impose the condition
\begin{equation}
|b_B| \lesssim 10^\circ \label{bc1}
\end{equation}
as a criterion for the preferred sky location of an extragalactic  burst. 
However, bursts located too close to the Galactic plane (e.g., \( |b_B| \lesssim 1^\circ \)) may be disfavored observationally due to interstellar absorption and source confusion, both of which can hinder counterpart identification. 

Using conditions~(\ref{theta_volume}), and (\ref{bc1}), together with  $\cos\theta_B=\cos l_B \cos b_B$, we obtain
\begin{equation}
0 \le l_B \lesssim 80^\circ \quad \text{or} \quad 280^\circ \lesssim l_B \le 360^\circ \label{bc2}
\end{equation}
for the preferred Galactic longitude of an extragalactic burst. 
\subsection{Width of a Search Ring}

The precision of the hybrid scheme is primarily limited by uncertainties in two quantities:  
\(r\), the distance to the Galactic Center, and \(R\), that to the extragalactic burst.
According to Eq.~(\ref{dgc}), the fractional uncertainty in \(r\) is approximately \(\Delta r / r \sim 0.005\).

In Eq.~(\ref{t2}), we compare the uncertainty in the first term with the magnitude of the second term.  
Although the second and higher-order terms can be evaluated using the measured distance to the burst (as in the case of SN1987A),  
our present comparison is intended to clarify the condition under which the plane wave approximation is valid in calculating for the angle $\beta$. 

We find that if
\[
\frac{\Delta r}{r} \gtrsim \frac{r}{R},
\]
then the second term becomes smaller than the uncertainty in the first term.  
This inequality can be rewritten as
\begin{equation}
R \gtrsim \frac{r^2}{\Delta r} \sim 2~\mathrm{Mpc}. \label{pwa}
\end{equation}

Therefore, given the current level of uncertainty in \(r\),  
the plane wave approximation can be reliably applied for reference bursts located beyond the Local Group.

For a sufficiently distant burst, Eq.~(\ref{t2}) yields the width of the search ring as
\begin{equation}
\Delta \beta \sim \frac{\beta}{2} \cdot \frac{\Delta r}{r}  \sim 0.6~\mathrm{arcmin} \left( \frac{\beta}{4^\circ} \right) \left( \frac{\Delta r / r}{5 \times 10^{-3}} \right). \label{DB}
\end{equation}

The search ring expands self-similarly over time.  
If this signaling scheme is indeed adopted by extraterrestrial transmitters, we can effectively compress  the search direction   by a factor of \(\Delta r / r\).  
It should be noted that this compression factor applies only to the search around an individual   reference burst.  
In fact, we can  make an implicit adjustment to commonly use a limited number of conspicuous bursts. If this really works, the total compression factor becomes significantly smaller than \(\Delta r / r\).

\begin{table*}[htbp]
\centering
\caption{Comparison of Conspicuous Explosive Events}
\label{tab:rep}
\begin{tabular}{lcccccccccc}
\toprule
Object &present  $\tau$ & RA [deg] & DEC [deg] & $l_B$ [deg] & $b_B$ [deg] & $R$ & $\theta_B$ [deg] & $r \cos \theta_B$ [kpc] & $\beta$ [deg] & time at $\beta = |b_B|$ \\
\midrule
SN1987A     & 38.2 yr & 83.9   & $-69.3$ & 279.7  & $-31.9$ & $52 \pm 5$ kpc      & 81.8 & 1.17 & 12.8 & 1987+290 \\
GW170817    & 7.8 yr  & 197.5  & $-23.4$ & 308.4  & +39.3    & $40^{+5}_{-4}$ Mpc  & 61.3 & 3.94 & 2.0  & 2017+3800 \\
GRB221009A  & 2.5 yr  & 288.3  & +19.8    & 53.0  & +4.3    & $\sim$700 Mpc       & 53.1 & 4.92 & 1.1  & 2022+46 \\
\bottomrule
\end{tabular}
\end{table*}

\section{Extragalactic Bursts }
In this section, we consider three types of explosive events, namely supernovae, NSB mergers, and GRBs, as potential extragalactic reference bursts.
For each category, the brightest event with reliable observational records is listed in Table~\ref{tab:rep}.  
Notably, all three representative events are located on the upstream side of the Galactic Center, as defined by Equation~(\ref{theta_downstream}).  
In the following, we examine these representative examples in greater detail.

\subsection{Supernovae}

SN 1987A, which occurred in the Large Magellanic Cloud (LMC), reached an apparent magnitude of approximately 2.9 and remains the brightest extragalactic supernova ever recorded.

In contrast, SN 1885A, the second brightest  extragalactic supernova, occurred near the center of the Andromeda Galaxy (M31), with a peak apparent magnitude  around 6 \citep{jones1976sandromedae}. It represents the oldest  reliably recorded extragalactic supernova and historically played an intriguing role in the Island Universe debate \citep{devaucouleurs1985sandromedae}.

According to \citet{tammann1994galactic}, the estimated supernova rate is about 0.6 per century in the LMC and the Small Magellanic Cloud (SMC), and approximately 1.8 per century in M31 and M33. These are  primary extragalactic supernova sites in our Local Group.

The LMC and SMC are located at approximately \((l, b) = (280^\circ, -33^\circ)\) and \((303^\circ, -44^\circ)\), respectively, placing them upstream from the Galactic Center satisfying  Eq.~(\ref{theta_downstream}). In contrast, M31 and M33 lie downstream, with Galactic longitudes of \(l=121^\circ\) and \(134^\circ\), respectively. Consequently, SN 1885A cannot be utilized in our scheme.

\begin{figure}[t]
 \begin{center}
  \includegraphics[width=8cm,clip]{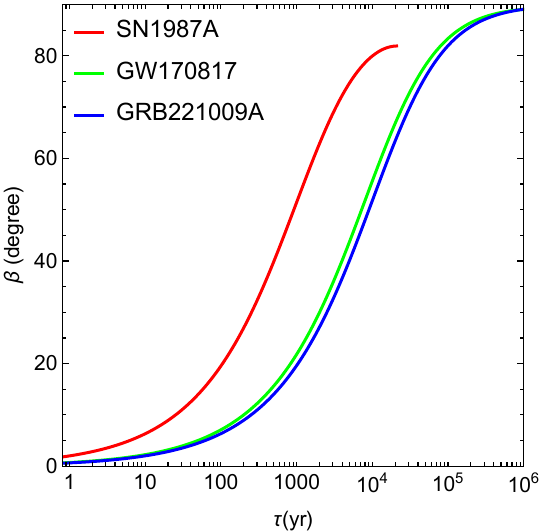}
   \caption{The opening angle $\beta$ for the search ring around the three representative extragalactic bursts. The horizontal axis shows the time $\tau$ since the observation of each burst. }
  \label{figure:fig4}
 \end{center}
\end{figure}

The distance to SN 1987A is still uncertain at the level of approximately 10\% (see e.g., \citealt{2023ApJ...949L...9C}), which dominates the width of the corresponding search ring. The angular separation from the Galactic Center is relatively large, \( \theta_B = 81.8^\circ \), and lies outside the range defined by Equation~(\ref{bc2}). As a result, the corresponding survey depth is shallow, and the opening angle becomes wide, \( \beta = 12.8^\circ \), even for a time delay of only about 40 years.
 The search ring reaches the Galactic midplane (\( b = 0 \)) after approximately 250 years.  In Figure~\ref{figure:fig4}, we present the opening angle \( \beta \) of the ring as a function of elapsed time \( \tau \) since 1987, using the full expression of Eq.~(\ref{bast}). 
 
{At the time difference \( \tau = 2.1\times10^{4}~\mathrm{yr} \),  
the opening angle \( \beta \) reaches its maximum value of \( 81.9^\circ \) (coincidentally close to \( \theta_B = 81.8^\circ \)), satisfying the tangential condition given in Eq.~(\ref{eq:tangent-condition}).  
Because of the relatively small distance parameter (\( R = 52~\mathrm{kpc} \)), the time profile of \( \beta \) deviates noticeably from the expression in Eq.~(\ref{tt3}) derived under the plane-wave approximation.  
In contrast, this approximation works well for the more distant bursts GW170817 and GRB~221009A discussed in the following two subsections. In Fig.~5, the two curves show a horizontal offset due to the difference in the  projection factor $\cos \theta_B$ (see Equations~(\ref{tt3}) and (\ref{tt4})).
}

\subsection{Neutron Star Binary Merger}

In the LVK dataset, GW170817 remains the only NSB merger with secure electromagnetic counterparts \citep{KAGRA:2021duu}. Its basic parameters, including a distance of \( R \sim 40\,\mathrm{Mpc} \) inferred from GW data, are listed in Table~\ref{tab:rep}. According to Eq.~(\ref{pwa}), the plane wave approximation is safely applicable in this case. The current opening angle of the search ring is approximately \( \beta \sim 2^\circ \), and it will take about 3800 years for the ring to reach the Galactic midplane.

Among other LVK events, GW190425 is also considered a candidate   NSB merger at a distance of  \(\sim160\,\mathrm{Mpc}\) \citep{LIGOScientific:2020aai}. However, no electromagnetic counterpart was identified, partly because its sky localization was more than two orders of magnitude worse than that of GW170817.

According to the LVK collaboration \citep{KAGRA:2021duu}, the estimated merger rate of NSBs is \(10\text{--}1400\,\mathrm{Gpc^{-3}\,yr^{-1}}\) (90\% credible interval). Using the lower end of this range, the expected occurrence rate of NSB mergers within a distance of \(40\,\mathrm{Mpc}\) is approximately one event every 400 years.

The planned space GW detector LISA will not be able to detect NSB beyond the local group \citep{LISA:2022yao}. A follow-on mission such as DECIGO, has the potential to observe an NSB at the distance of $R\sim100$Mpc at $\sim10$ yr before the merger \citep{Seto:2001qf,Kawamura:2020pcg}.  In Fig.~3, a civilization with such a detector can transmit from upstream of the red burst surface, modestly increasing the signaling region described in Sec.~5.1.

\subsection{Gamma-ray Bursts}

GRB~221009A attracted considerable attention due to its exceptionally high fluence \citep{2023ApJ...946L..31B,2023ApJ...952L..42L,2023ApJ...949L...7F}.
The redshift of this burst was measured to be \( z = 0.15 \), placing it at a relatively nearby cosmological distance for a GRB. Its isotropic-equivalent energy, \( E_{\rm iso} \gtrsim 10^{55} \) erg, was extraordinarily large, likely a result of a highly collimated jet with a narrow opening angle.

In Table~\ref{tab:fluences}, we list seven GRBs with the highest observed fluences, based on data from \citet{2023ApJ...946L..31B}. The original list (Table~2 in their paper) included 13 GRBs, but here we focus on those with identified host galaxies. The fluence of GRB~221009A is roughly 40 times higher than that of the second-brightest event, GRB~230307A. Statistical estimates suggest that such an event may be detected at the Sun only once every 10,000 years \citep{2023ApJ...946L..31B}. Owing to its extreme rarity and exceptional luminosity, GRB~221009A has been informally referred to as the ``Brightest Of All Time'' (BOAT) GRB.

We also added the Galactic coordinates \((l_B, b_B)\) for each GRB in Table~\ref{tab:fluences}. Only the BOAT GRB satisfies both conditions~(\ref{bc1}) and~(\ref{bc2}) for an ideal sky position. In particular, its low Galactic latitude of \( b_B = 4.3^\circ \) makes it especially suitable for efficiently scanning Galactic civilizations. Taking into account the \( \sim10\% \) sky fraction of such preferred directions, the detection rate of similarly favorable GRBs may be as low as once per 100{,}000 years.

The combined characteristics (its exceptionally high fluence and favorable position near the Galactic plane) establish GRB~221009A as a compelling reference  candidate for our hybrid scheme.

\begin{table*}[htbp]
\centering
\caption{GRBs with high observed fluence}
\begin{tabular}{lcccccc}
\toprule
GRB Name & Fluence (erg cm$^{-2}$) & Redshift ($z$) & RA [deg] & DEC [deg] & $l_B$ [deg] & $b_B$ [deg] \\
\midrule
GRB 221009A  & 0.21               & 0.151  & 288.3  & +19.8  & 53.0   & +4.3   \\
GRB 230307A  & $4.56 \times 10^{-3}$ & 0.065  & 60.9   & -75.4  & 289.5  & -36.3  \\
GRB 130427A  & $2.86 \times 10^{-3}$ & 0.34   & 173.1  & +27.7  & 206.5  & +72.5  \\
GRB 160625B  & $1.23 \times 10^{-3}$ & 1.406  & 308.6  & +6.9   & 51.5   & -18.9  \\
GRB 160821A  & $1.17 \times 10^{-3}$ & 0.16   & 171.2  & +42.3  & 166.5  & +66.7  \\
GRB 180914B  & $1.21 \times 10^{-3}$ & 1.096  & 332.4  & +24.7  & 82.3   & -25.1  \\
GRB 140219A  & $1.20 \times 10^{-3}$ & 0.12   & 156.0  & +6.5   & 236.7  & +49.4  \\
\bottomrule
\end{tabular}\label{tab:fluences}
\end{table*}

\section{Discussion }
In this section, we discuss practical issues related to the proposed hybrid scheme.

\subsection{Potential Impact of Echoes}

Electromagnetic echoes have been observed in association with several astrophysical transients, such as SN~1987A \citep{2023ApJ...949L...9C} and GRB~221009A \citep{2023ApJ...946L..30T}, where scattered lights appeared near  the explosion site on the sky. Although  intentional signals would be  fundamentally different  from such echoes,  the latter might introduce  background emission that raises the effective noise level in SETI observations.

{
At a given time $\tau$ and  for the corresponding search direction [with  opening angle $\beta$ determined through Eq. (\ref{bast})], the position of the scattering material responsible for such echoes is uniquely determined to be point $K$ in Fig.~3 (as discussed in subsection~5.1; see also \citealt{2023ApJ...949L...9C}).
}
Depending on the actual observational conditions, it may be necessary to exclude portions of the search ring that are unfavorable. In general, scattering material tends to be more abundant toward the Galactic disk, where the stellar density is also higher.

\subsection{Effect of Aberration}

Objects near the Galactic plane exhibit motion primarily due to Galactic rotation, with typical velocities on the order of \(200\,\mathrm{km\,s^{-1}}\). When considering signaling on Galactic scales, such motion can alter the apparent directions of  signal propagation through the  aberration effect.

This effect is described within the framework of special relativity. When an observer moves with velocity \(\vec{v}\), the apparent direction of incident light  is shifted on the sky. In the limit of small velocities, the angular shift \(\Delta\theta\) is approximately given by \(\Delta\theta \sim v_\perp / c\), where \(v_\perp\) is the component of the velocity perpendicular to the line of sight \citep[see, e.g.,][]{LLII}.

To evaluate the aberration effect in our context, it is necessary to define a reference frame. On Galactic scales, the rest frame relative to the Galactic Center serves as a natural common choice from the perspective of a Schelling point.

In contrast, our SETI observations are conducted in a frame moving with respect the Galactic Center.  If the relative velocity is \(v \sim 200\,\mathrm{km\,s^{-1}}\) and the opening angle of the search ring is \(\beta (\ll 1)\), the ring may be deformed relative to the burst direction by at most \(\sim \beta v/c\). However, this shift is smaller than the angular width \(\delta \beta\) given by Eq.~(\ref{DB}) due to the uncertainty in the distance   to the Galactic Center. Therefore, at current precision levels, the effect of aberration can be safely ignored.

\subsection{Number of Extragalactic References}
In my previous studies, I considered using a relatively small number (typically fewer than ten) of Galactic references, such as historical supernovae or  short-period NSBs. In that approach, each  search ring sweeps across a large portion of the sky in a timescale of \( r / c \sim 10^4 \) years. As a result, only a limited portion of the sky can be probed within a century.

However, the typical lifetime of a highly advanced civilization might be shorter than a few hundred years (see e.g., \citealt{1993Natur.363..315G}). In such cases, long-timescale search strategies may risk missing potential signal detection opportunities. An alternative approach is to make use of a larger number of extragalactic bursts in parallel, enabling signal transmission and reception over a wide range of the Galactic plane within a relatively short period of time. This highlights a potential trade-off between the conspicuousness of individual bursts and the achievable sky coverage within a limited observational timeframe.

\subsection{Evolving Candidates for  References}

In this subsection, we discuss the potential extension of  the hybrid scheme.

When considering signal transmission or detection on Galactic scales, the Galactic Center appears to be a natural choice as a spatial reference, particularly from a Schelling point perspective. However, in cases where signaling is limited to a relatively local region near the Sun, it may be worth considering alternative positional references. Such objects would need to be distinctive, and their distances would need to be estimated with high precision.

In  the near future,  new types of astronomical burst will emerge as viable candidates for extragalactic reference bursts. Among these,  the mergers of massive black holes stand out due to their extremely large energy release \citep{LISA:2022yao}. 
 The key issue is the angular localization of such events. GW data alone are unlikely to provide the sub-minute directional accuracy required by the present scheme. However, if the mergers are accompanied by observable electromagnetic transients, this limitation could be overcome, and they might become strong candidates for temporal markers.

These dual implementations would extend the applicability of the hybrid  framework   and enhance its potential relevance to future SETI strategies.

\section{Summary}

This work builds on my previous studies of the \textit{concurrent signaling scheme}, a concept in which both transmitter and receiver refer to a common astrophysical event to coordinate the timing and direction of interstellar communication. Rooted in Schelling point logic, this approach has the potential to substantially reduce the parameter space that SETI must explore. Unfortunately, the proposed Galactic anchors come with significant drawbacks that limit the effectiveness of the concurrent signaling scheme.

To address these challenges, I have proposed a new hybrid strategy that combines two elements: (i) the Galactic Center as a geometrical reference point with a well-determined distance, and (ii) an extragalactic conspicuous burst to introduce a temporal marker. This combination enables a substantial reduction of the target sky directions by a factor of at least  200, based solely on existing astronomical data.

As candidates for the extragalactic burst, I examined three categories: supernovae,  NSB mergers, and GRBs. For each category, I evaluated the most conspicuous events and their suitability as temporal markers. Among these, GRB~221009A is particularly promising   due to its exceptionally high fluence and favorable sky position near the Galactic plane. Such advantageous GRBs might become available at the Sun's location only once every 100,000 years.

\vspace{5mm}
\nolinenumbers
The author is grateful to the anonymous referee for many valuable comments that helped improve the manuscript.
\bibliography{ref2}

\end{document}